\documentclass[
prb,
longbibliography,
superscriptaddress,
twocolumn,
 amsmath,amssymb,
]{revtex4-2}

\usepackage{graphicx}% Include figure files
\usepackage{dcolumn}% Align table columns on decimal point
\usepackage{bm}% bold math
\begin{document}

%\preprint{APS/123-QED}

\title{Low energy electrodynamics and a hidden Fermi liquid in the heavy-fermion CeCoIn$_5$}

\author{L.Y. Shi}
 \affiliation{William H. Miller III Department of Physics and Astronomy, Johns Hopkins University, Baltimore MD, USA}

\author{Zhenisbek Tagay}
 \affiliation{William H. Miller III Department of Physics and Astronomy, Johns Hopkins University, Baltimore MD, USA}

 \author{Jiahao Liang}
 \affiliation{William H. Miller III Department of Physics and Astronomy, Johns Hopkins University, Baltimore MD, USA}
 
 \author{Khoan Duong}
 \affiliation{Department of Physics, Cornell University, Ithaca, New York 14853, USA}

 \author{Yi Wu}
 \affiliation{Department of Physics, Cornell University, Ithaca, New York 14853, USA}

 \author{F. Ronning}
 \affiliation{Institute for Materials Science, Los Alamos National Laboratory, Los Alamos, New Mexico, 87545, USA}

 \author{Darrell G. Schlom}
 \affiliation{Department of Materials Science and Engineering,
Cornell University, Ithaca, New York 14853, USA}
 \affiliation{Leibniz-Institut f{\"u}r Kristallz{\"u}chtung, 12489 Berlin, Germany}
  \affiliation{Kavli Institute at Cornell for Nanoscale Science, Ithaca, New York 14853, USA}
 
 \author{K.M. Shen}
 \affiliation{Department of Physics, Cornell University, Ithaca, New York 14853, USA}
 \affiliation{Kavli Institute at Cornell for Nanoscale Science, Ithaca, New York 14853, USA}
 
\author{N.P. Armitage}%
 \email{npa@jhu.edu}
 \affiliation{William H. Miller III Department of Physics and Astronomy, Johns Hopkins University, Baltimore MD, USA}
  \affiliation{Canadian Institute for Advanced Research, Toronto, Ontario M5G 1Z8, Canada}

\date{\today}

\begin{abstract}
We present time-domain THz spectroscopy of thin films of the heavy-fermion superconductor CeCoIn$_5$.  Below the $\approx$ 40 K Kondo coherence temperature, a narrow Drude-like peak forms, as the result of the $f$ orbital - conduction electron hybridization and the formation of the heavy-fermion state.  The complex optical conductivity is analyzed through a Drude model and extended Drude model analysis.  Via the extended Drude model analysis, we measure the frequency-dependent scattering rate ($1/  \tau $) and effective mass ($m^*/m_b$). This scattering rate shows a linear dependence on temperature, which matches the dependence of the resistivity as expected. Nevertheless, the width of the low-frequency Drude peak itself that is set by the {\it renormalized} quasiparticle scattering rate ($1 / \tau^* = m_b/ m^* \tau $) shows a $T^2$ dependence.   This is the scattering rate that characterizes the relaxation time of the renormalized quasiparticles.  This gives evidence for Fermi liquid state, which in conventional transport experiments is hidden by the strong temperature dependent mass.

\end{abstract}

\maketitle

%\tableofcontents
 
\section{\label{sec:level1} Introduction} 
 Heavy-fermion materials are canonical strongly correlated systems~\cite{RevModPhys.71.687}. A typical heavy-fermion material contains $f$ orbitals and hosts local moments as well as itinerant conduction electrons. At high temperatures, such systems show weakly correlated normal-metal behavior, while at low temperatures, correlations dominate. In the conventional picture two different interactions that compete with each other. The Kondo interaction describes the tendency for $f$ moments and itinerant carriers to form coupled singlets~\cite{RevModPhys.56.755,hewson1997kondo}. The Ruderman, Kittel, Kasuya, Yosida (RKKY) interaction is the interaction between local moments mediated by the itinerant electrons~\cite{https://doi.org/10.1002/pssb.201300005,PhysRevB.55.8064,PhysRevB.58.3584}.  When the Kondo interaction is dominant, the $f$ electrons couple to the itinerant carriers forming a Kondo singlet at low temperatures.  Despite strong interactions, this hybridized state is frequently describable in terms of Fermi-liquid (FL) theory, albeit one with a terrifically enhanced quasiparticle mass.  In lattice systems the formation of this collective state typically occurs at a lower temperature scale $T^*$ than that of the single-ion Kondo interaction.  Some heavy-fermion systems have been observed to deviate from the FL behavior in their electrical resistivity and specific heat~\cite{PhysRevLett.72.3262,PhysRevLett.112.206403}. One explanation of these non-FL properties is the proximity to a nearby quantum critical point~\cite{PhysRevLett.67.2886,PhysRevB.74.132408,doi:10.1143/JPSJ.81.011002}.

CeCoIn$_5$ is a widely studied heavy-fermion material.  It belongs to the CeMIn$_5$ (M = Co, In, Rh) material class, which has attracted widespread attention for their unconventional superconductivity and proximate quantum critical point~\cite{Petrovic2001L337,Sarrao2007,PhysRevB.96.045107, Kirchner2020,PhysRevLett.89.157004,PhysRevB.80.064512}. CeCoIn$_5$ has the highest superconducting transiton temperature of the group with a superconducting temperature $T_c$ = 2.3 K. Below the Kondo lattice coherence temperature $T^* \approx 40$ K and above $T_c$, it shows signatures of a non-FL metal including $T$-linear resistivity~\cite{Paglinoe2007}.  The strong correlations in the heavy-fermion state typically manifest in the infrared spectrum~\cite{Basov2011471,Chen_2016, PhysRevB.85.155105, PhysRevB.93.085104}. In CeCoIn$_5$, the $f$ electron dynamics, particularly the hybridization gap that reflects the interaction between conduction and $f$ electrons, have been studied by infrared spectroscopy ~\cite{PhysRevB.75.054523,PhysRevB.72.045119, PhysRevB.65.161101, Okamura_2015,LEE2023106376,PhysRevLett.124.057404}. Nonetheless, the energy scale of the heavy-carriers dynamics is typically located at even lower energies in the THz range.  This has been only studied by low frequency resolution quasi-optical measurements~\cite{ Scheffler2013,PRACHT201631}. The low-energy dynamics of the heavy electrons in the THz range is important for understanding signatures of non-FL behavior and possible quantum criticality in this system. The quality of these previous data was not sufficient to resolve detailed line shapes, but nevertheless could show a strong temperature dependence of the scattering rate and effective mass from a  Drude model analysis. At low temperatures, a deviation from the simple Drude theory was observed, indicating its limitations in describing the charge dynamics in the heavy-fermion state. A sophisticated analysis based on a high quality optical spectral data would be important to reveal the behavior in this system.

In this work we use time-domain THz spectroscopy (TDTS) to study the low frequency complex conductivity of CeCoIn$_5$ films. The frequency-dependent optical scattering rate and effective mass were calculated using an analysis of the spectral weight and line shape. A strong enhancement of the effective mass was found that is consistent with the formation of the heavy-fermion state. Both the optical scattering rate and resistivity at low temperature show a linear dependence on temperature that has been taken to be evidence for non-FL behavior. However, the complex THz optical conductivity allows us to characterize the {\it quasiparticle} scattering, which manifests as the width of the low-frequency Drude peak.  Notably this quasiparticle scattering rate shows a $T^2$ dependence, giving evidence for a hidden Fermi liquid.  The presence of a Fermi liquid state is typically obscured in the resistivity by the strong temperature dependent mass renormalizations.

\section{\label{sec:level2} Experiments} 
CeCoIn$_5$ thin films were grown on 10 mm $\times$ 10 mm  MgF$_2$ substrates by molecular-beam epitaxy (MBE). Samples were grown in a Veeco GEN10 system with elemental sources of cerium (99.99$\%$), cobalt (99.99$\%$) and indium (99.99$\%$) in a background pressure better than $5 \times 10^{-9}$ torr, following a similar procedure detailed by Mizukami et al.~\cite{Mizukami_2011}. The elemental fluxes were calibrated using a quartz crystal microbalance and x-ray reflectivity measurements of film thicknesses before growth. The MgF$_2$ substrates were first pre-annealed at 750$^\circ$  before growth until a clear reflection high-energy electron diffraction (RHEED) pattern was observed, and then lowered to 375$^\circ$  for the film deposition. The growth rate was 1.5-2.5 nm/min and monitored in real-time using RHEED. Samples were characterized post-growth using x-ray diffraction and electrical resistivity measurements. 

The thickness of the films were limited at 100 nm to have sufficient THz transmission.  Residual resistivity ratios of approximately 5 are typical.  The sample was kept in a vacuum or helium environment after growth and before measurement to minimize oxidization. A bare MgF$_2$ substrate was used as a reference in the THz measurement.  The THz spectrum from 0.2 - 1.8 THz was measured on a modified fiber-coupled Toptica time-domain THz spectrometer. By dividing the Fourier transform of E field transmission through the sample by the Fourier transform of transmission through the bare MgF$_2$ reference substrate, the complex transmission $T(\omega)$ is obtained and the complex optical conductivity ($\sigma = \sigma _1 + i \sigma _2$) can be calculated from $T(\omega)$ by $T(\omega) = \frac{1+n}{1+n+\sigma d Z_0}e^{\frac{i\omega \Delta L(n-1)}{c}}$.  Here $n$ is the substrate index of refraction, $d$ is the film thickness, $\Delta L$ is a correction factor that accounts for thickness differences between the reference and sample substrates, and Z$_0$ is the impedance of free space (377 $\Omega$). 

\begin{figure}[htbp]
	\centering
	\includegraphics[width=7cm]{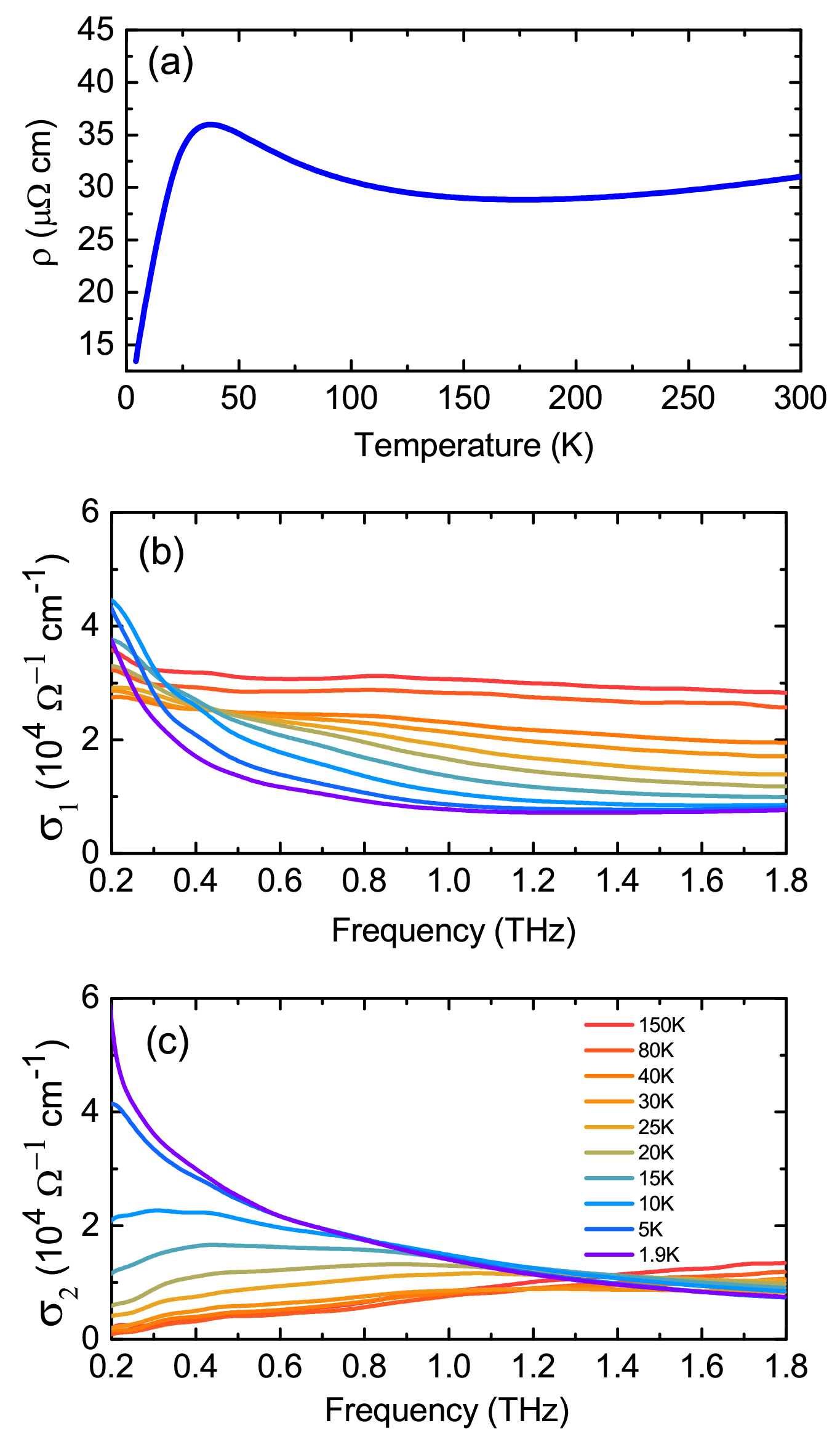}\\
	\caption{(a) dc resistivity as a function of temperature for the CeCoIn$_5$ thin film. (b) Real part of the optical conductivity ($\sigma _1$) at different temperatures. (c) Imaginary part of the optical conductivity ($\sigma _2$) at different temperatures.}\label{Fig:1}
\end{figure}

\section{\label{sec:level3} Experimental results} 
Fig. \ref{Fig:1}(a) shows the dc resistivity of the thin film from 5 K to 300 K measured in a van der Pauw geometry. A small correction to account for inaccuracies in the geometrical factor is applied by normalizing it to the THz conductivity at 150 K (where there is little frequency dependence).  All the previously reported features of CeCoIn$_5$ are clearly observed, indicating the good quality of the film. Starting from 300 K, the resistivity shows a conventional metal behavior that first decreases with the temperature and goes to a minimum around 150 - 170 K.  Below 150 K, the resistivity increases again due to incoherent Kondo scattering off of localized $f$ electrons and yields a maximum at $T^*$ = 40 K.   This maximum is taken at the onset temperature of the coherent Kondo state as at lower temperatures the resistivity drops quickly. The superconducting transition temperature of thin films is expected at 2 K, which is at the edge of our THz measurement temperature range. 

\begin{figure}[htbp]
	\centering
	\includegraphics[width=8.5cm]{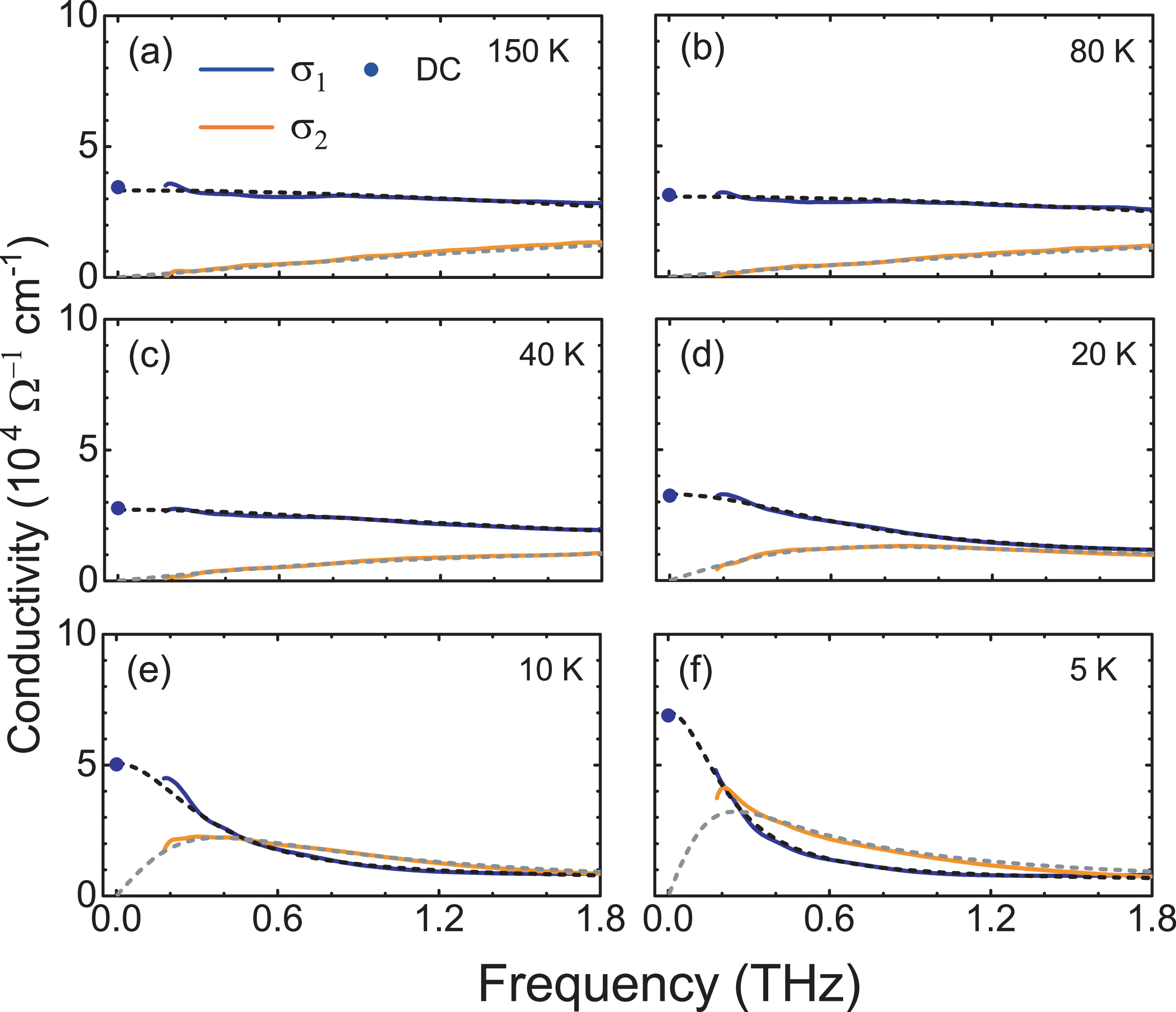}\\
	\caption{(a) - (f) Real and imaginary parts of the optical conductivity spectrum at six temperatures. Solid lines indicate experimental data, dashed lines show the two term Drude model fit, and solid circles show the dc conductivity.}\label{Fig:2}
\end{figure}

Figs. \ref{Fig:1}(b) and 1(c) show the real and imaginary parts of the THz complex conductivity $\sigma$ at different temperatures. At 150 K, the real part is flat and the imaginary part is small consistent with strong scattering. Until 50 K, the conductivity remains largely of the same shape with its overall scale increasing (consistent with the dc data). Below 50 K, the spectral weight shifts toward low frequencies and forms a narrow Drude-like peak, reflecting the coupling of $f$ and conduction electrons, which enhances the near Fermi energy density of states~\cite{doi:10.1073/pnas.2001778117}. We use a Drude model to fit the real and imaginary parts of the THz conductivity simultaneously with the dc conductivity. The fits at several temperatures are shown in Fig. \ref{Fig:2}. At high temperatures, the spectrum can be well fit by a single zero-frequency oscillator with a broad width. Below 50 K, the fitting requires at least two zero-frequency oscillators, one broad for the flat spectrum extended to beyond the measurement range and one narrow for the renormalized heavy fermion mode. The fits well match the dc data and both complex components of the THz response.   It is important to note that the two-Drude-component fit does not necessarily correspond to two distinct charge carrier species. It is simply a minimal Kramers-Kronig consistent parametrization of the THz and dc data.  This is not to say that such a fit does not have physical content.   In the case where there is a large peak that carries most of the spectral weight, then its width is the actual relaxation rate of this portion of the current.  We should also note that the 1.9 K data are below the superconducting transition temperature (2 K) of the CeCoIn$_5$ thin film and unsurprisingly the spectrum at 1.9 K deviates from the temperature trends at the lowest frequencies, as shown in Figs. \ref{Fig:1}(b) and 1(c). The real part of the optical conductivity decreases while the imaginary part increases and could not be fit without adding additional terms to the fitting routine\cite{PhysRevLett.89.157004, Zhou2013, PhysRevLett.130.076301}. The electrodynamics of the superconducting state will be pursued in a future work.

To further analyze the heavy-carrier dynamics, we use the extended Drude model to monitor two important quantities that give rise to this phenomenoma: the frequency dependent scattering rate $1/ {\tau (\omega)}$ and the normalized effective mass $m^* (\omega)/m_b$:
\begin{align}
\dfrac{1}{\tau (\omega)} = \dfrac{\omega ^2 _p}{4 \pi  } \mathrm{Re} \Big [\dfrac{1}{\sigma (\omega)} \Big],  \; \; \; \omega 
     \dfrac{m^* (\omega)}{m_b} = - \dfrac{\omega ^2 _p}{4 \pi } \mathrm{Im} \Big[\dfrac{1}{\sigma (\omega)}\Big] \label{eq3}
 \end{align}
Here $\omega_p$ is the plasma frequency and $m_b$ is the ordinary electron band mass.  Although the extended Drude model might appear to require the existence of a single quasi-particle band, the quantities extracted are more general than that.   Eqs.~\ref{eq3} are proportional to the real and imaginary parts of the frequency dependent {\it resistivity} and express the rates of relaxation and inertial response of the macroscopic current.  Corrections to the extended Drude analysis that might have come from subtracting an $\epsilon_\infty$ from $\sigma$ that arises from the finite polarizability of the lattice are negligible.

The value of the scattering rate and effective mass depends on the value of the plasma frequency $\omega _p$, which is a measure of the total Drude spectral weight of all free charge carriers.   One may get it via integration of the spectra up to high energies from the partial sum rule $\omega_p^2/8 = \int_0^\Omega \sigma_1(\omega) d\omega$.  As a practical matter it can be hard to separate the free-carrier spectral weight from other excitations (e.g., to choose an appropriate $\Omega$). Although the uncertainty of  $\omega _p ^2$ brings uncertainty in the absolute value of these quantities, it does not affect their frequency or temperature dependence as $\omega _p ^2$ is a T and $\omega$ independent prefactor. Here we use an estimated value from a previous report~\cite{LEE2023106376}, $\omega _p$ = 29,700 cm$^{-1}$ (892 THz) of CeCoIn$_5$.  $m^*/m_b$ has the straightforward interpretation of the ratio of the mass enhancement of the quasiparticles to the uncorrelated band mass.   In a limit where vertex corrections can be neglected, $1/\tau(\omega)$ is a quantity that can be related to {\it electronic} self-energy e.g., $1/\tau = -2 \mathrm{Im} \Sigma(\bf{k_F},\omega) $.  Note that this quantity does not give the width of the Drude peak directly (which is the rate that charge currents actually decay).   The width of the Drude peak is the fully renormalized quasiparticle scattering rate ($1 / \tau^* = \frac{m_b}{m^*} \frac{1}{\tau} $), which will be discussed further below.   Similar to the above two Drude component analysis, quite independent of any microscopic perspective the extended Drude model has significance in that in characterizes how fast a particular frequency component of a charge current relaxes.

\begin{figure}[htbp]
	\centering
	\includegraphics[width=8.5cm]{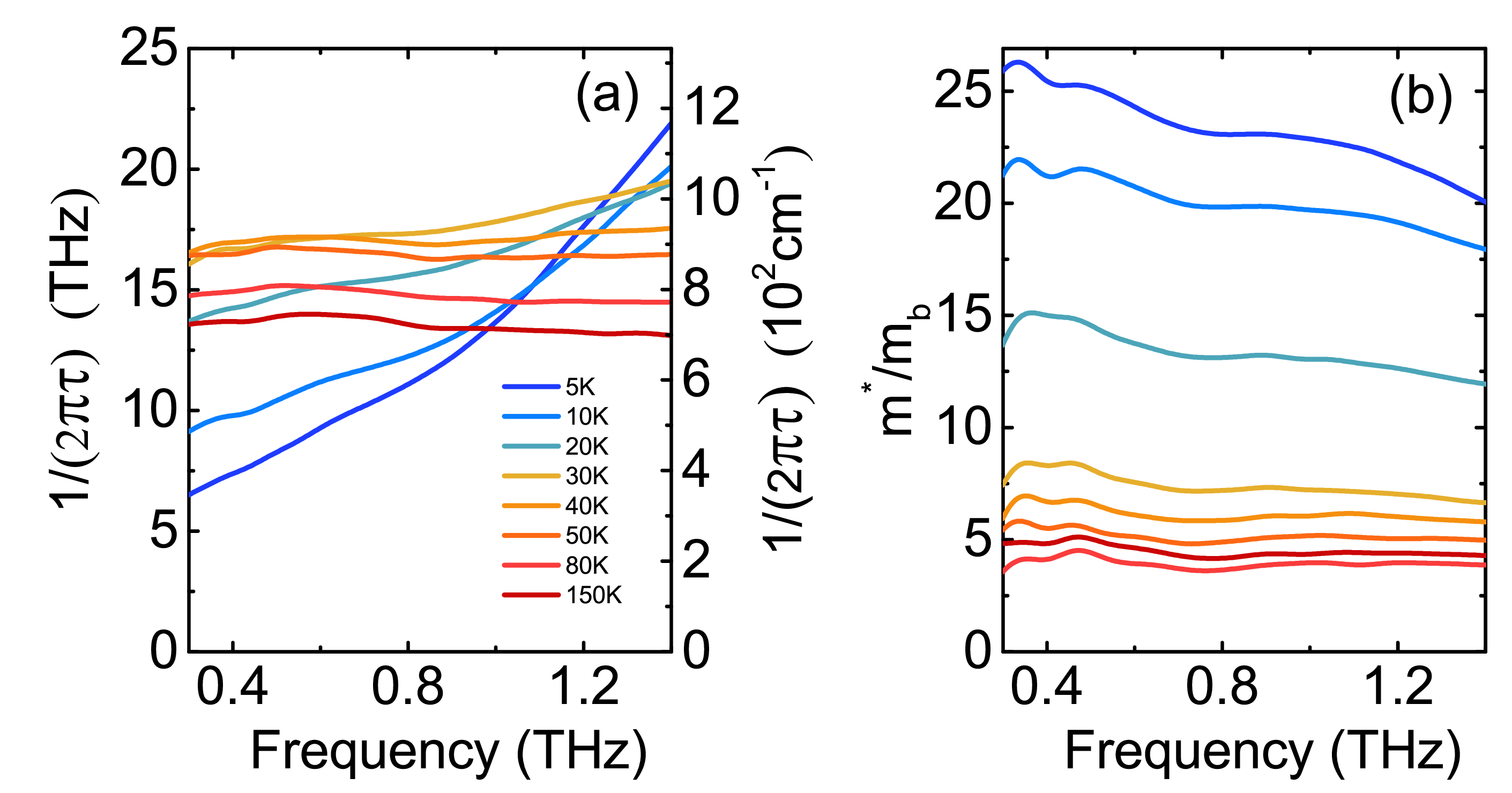}\\
	\caption{Scattering rate and effective mass derived from the extended Drude model. (a) The scattering rate as a function of frequency at selected temperatures. (b) The effective mass as a function of frequency at the same temperatures. }\label{Fig:3}
\end{figure}

Fig. \ref{Fig:3} shows the frequency dependence of the scattering rate $1/ {\tau (\omega)}$ and effective mass $m^* (\omega)/m_b$ at select temperatures from 5 K to 150 K, calculated from Eq. \ref{eq3}. As shown in Fig. \ref{Fig:3}(a), at high temperatures, the scattering rate is almost constant over the THz range.   This is consistent with the fact that the conductivity in this range is fit well with a single Drude oscillator. The scattering rate first gradually increases when the temperature decreases, in correspondence with the dc resistivity and increased scattering off the local $f$ moments. Below $T^*$, the scattering rate becomes strongly frequency-dependent indicating the formation of the low temperature heavy-fermion state. The optical mass is flat in frequency and small at high temperature, but becomes strongly frequency and temperature dependent at low temperature.  As shown in Fig. \ref{Fig:3}(b), the effective mass is enhanced by a factor of 25 over the band mass at the low temperature of 5 K. 

\begin{figure}[htbp]
	\centering
	\includegraphics[width=8.5cm]{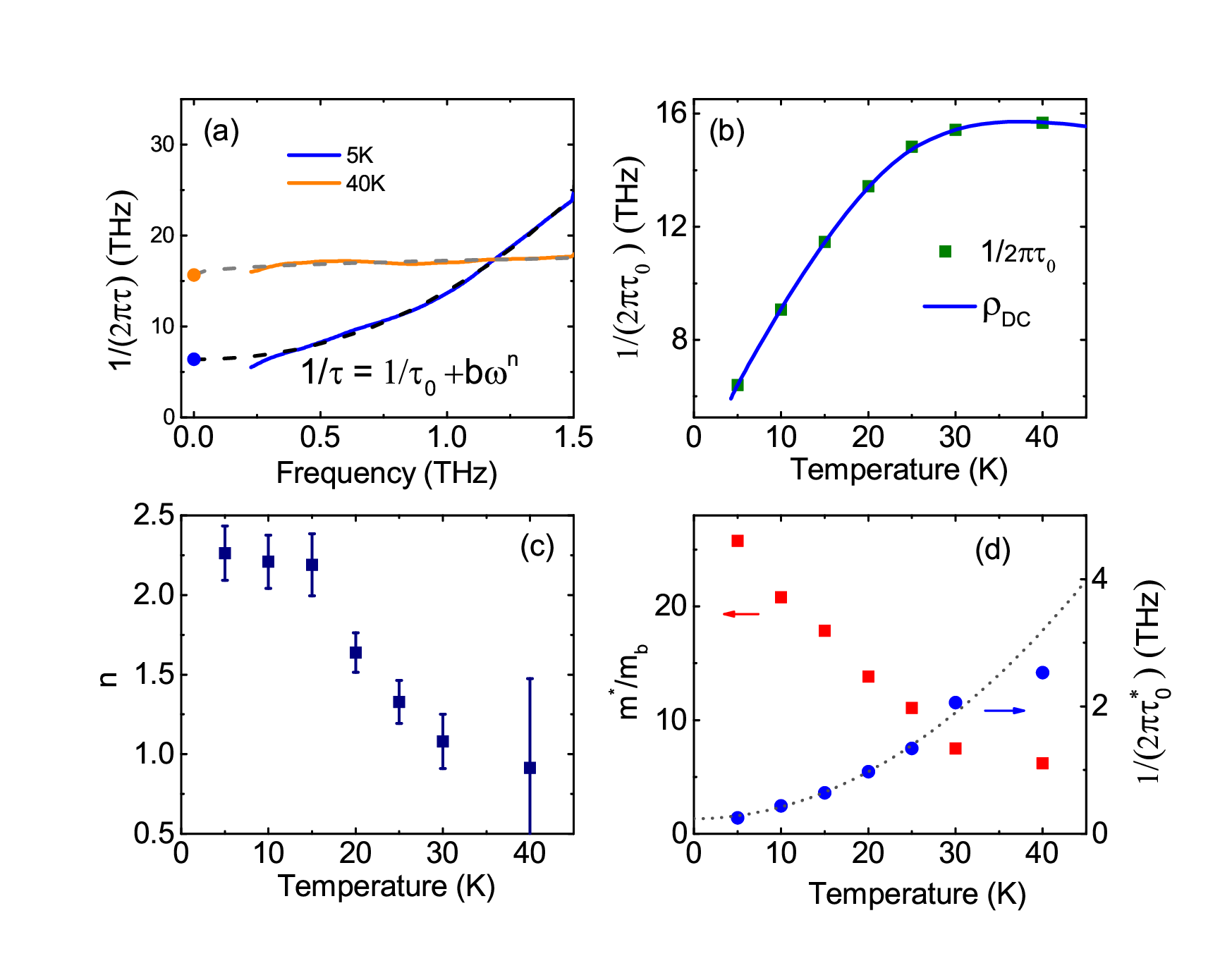}\\
	\caption{(a) Zero-frequency scattering rates and power law fits. The solid lines show experimental data. The dc data are shown as solid circle. The dashed lines are the power law fits $\frac{1}{\tau (\omega , T)} = \frac{1}{\tau (0, T)} + b \omega ^n$. (b) The green squares are the zero-frequency scattering rate, calculated at each temperature by inverting the two-Drude model fit of the conductivity. The blue line is the dc resistivity scaled by $\omega_p ^2 / 4 \pi$ to match the scattering rate. (c)  Power of the frequency dependence of $\frac{1}{\tau(\omega)}$ at different temperatures. (d) Red squares are the effective mass at zero frequency as a function of temperature. The blue circles show $1/ \tau ^*$, the scattering rate normalized by the mass enhancement. The dotted line shows a $T^2$ fit.}\label{Fig:4}
\end{figure}

At low temperatures when the incoherent scattering from local moments is suppressed in favor of a coherent heavy-fermion state, the frequency and temperature dependent scattering arises mainly from electron-electron interactions. FL theory predicts a scattering rate $\frac{1}{\tau (\omega , T)} = \frac{1}{\tau'} + A[(\hbar \omega)^2 + (2 \pi k _B T)^2]$~\cite{Maslov_2017,PhysRevB.86.155137,PhysRevB.87.115109}. $\frac{1}{\tau'} $ is the small temperature and frequency independent residual impurity scattering.  We can characterize the zero-frequency scattering rate by inverting our two-Drude component fit of the conductivity, which provides a natural way to connect THz to the dc data.  We then fit the frequency-dependent scattering to a form $\frac{1}{\tau (\omega , T)} = \frac{1}{\tau (0, T)} + b \omega ^n$, where $\frac{1}{\tau (0, T)}$ is the zero-frequency scattering rate and $n$ is a temperature-dependent exponent.  In performing these fits, we first fixed the zero-frequency scattering from the inversion of the two Drude fits and then find the parameters of the power law. The fits from 0.25 THz to 1.5 THz at 5 K and 40 K are shown in Fig.~\ref{Fig:4}(a).  Fig.~\ref{Fig:4}(b) shows the temperature dependence of the zero-frequency scattering rate overlaid with the dc resistivity (normalized by $\omega_p ^2 / 4 \pi$).  Both follow a linear behavior at low temperature, which notably deviates from a $T^2$ FL behavior. Fig. \ref{Fig:4} (c) shows the exponent $n$ of the fit at different temperatures. The resulting exponent is influenced by the upper limit of the fitting range. We set an error bar by noting the deviation of $n$ by changing the upper limit from 1.5 THz to 0.8 THz. In contrast to the linear dependence on temperature, the exponent of frequency dependence increased to around 2 at low temperature. This dependence is consistent with a Fermi liquid, but raises questions as to how it can be reconciled with the linear temperature dependence of $1/\tau$ and $\rho$?

By using the same extended Drude model analysis and an  inversion of the two Drude fit, we can also find the zero-frequency effective mass $m^* / m_b$ at each temperature, as shown in Fig. \ref{Fig:4}(d). This ratio of the effective mass is around 5 above 50 K, starts to be enhanced below 40 K, and rises to 25 at 5 K. As mentioned above, the absolute value of the effective mass is set by the value we use for the plasma frequency, but the temperature dependence of the effective mass enhancement is clear.

\section{\label{sec:level4} Discussion} 

The dichotomy between the linear in $T$ and $\omega^2$ dependence of $1/\tau$ is notable.  In this regards, an important quantity to consider is the renormalized quasiparticle scattering rate $ \frac{1}{ \tau^* } = \frac{m_b}{m^*} \frac{1}{\tau} $.  A few comments are in order.  The normal Drude expression for the dc conductivity is proportional to $\tau/m_b$ and represents the conventional FL result that by some measure the mass renormalizations do not enter explicitly in the transport  directly~\cite{prange1964transport,varma1985phenomenological} and the current is controlled by the formal quantity $ \frac{1}{ \tau}$.   Indeed we found that the resistivity tracks the zero-frequency value of $1/\tau$ (Fig. \ref{Fig:4}(b)) with a temperature independent mass.  However if one considers that the current is carried by quasiparticles of mass $m^*$ and not carriers of mass $m_b$, then for consistency with the usual expression they must have a  scattering time $\tau^*$ e.g., $\sigma_{dc} \propto  \tau/m_b \propto  \tau^*/m^*   $    If the frequency dependence of the effective mass is weak at low energies then $1 / 2\pi \tau^*$ corresponds to the width of the Drude response at low frequency and is the rate of relaxation of a current of particles of mass $m^*$.

As noted in Refs.~\cite{xu2013hidden,deng2013bad,PhysRevLett.113.246404,yang2014optical}, many strongly correlated metals show (in both optics and numerics) show anomalous power laws in $T$ dependence of the resistivity.  In contrast, their fully renormalized scattering rates shows a quadratic $T$ dependence to high temperatures.  It was proposed that such systems manifest a {\it hidden} Fermi liquid state in which the $1/\tau^*$ scattering of quasiparticles goes as $T^2$ or $\omega^2$ at low energies, but that this behavior in the resistivity is masked by a strong temperature dependence of the effective mass, which onsets at much lower temperatures.  We can similarly calculate the normalized quasiparticle scattering rate $\frac{1}{\tau ^*} = \frac{m^*}{m_b} \frac{1}{\tau}$, as shown in Fig. \ref{Fig:4}(d), where we find that the resulting quasiparticle scattering rate can be fitted by a $T^2$ dependence.  In this regard, despite the many unconventional properties of CeCoIn$_5$ it appears to have quasiparticles which are subject to the usual Fermi-liquid like phase space constraints and therefore may be a ``hidden" Fermi liquid.  Note that this terminology of a ``hidden" Fermi liquid was also used by P.W. Anderson in a very different context~\cite{anderson2010hidden}, where he proposed that the physics of the cuprate superconductors in the superconducting state was related to that of a normal metal Fermi liquid in a physically inaccessible space, but one that the normal state of cuprates would have had $if$ they had the same kind of scattering as the superconducting state.  As he explained, ``In the ‘normal’ state, the strange metal, the wavefunction renormalization connecting the two, Z, is zero. When the gap opens, Z becomes finite; there is a coherent quasiparticle."~\cite{anderson2010hidden}.  This is obviously an almost unrelated usage of the terminology.

%%%  NPA \cite{yang2014optical}

\section{\label{sec:level5} Conclusion}

In summary, we used time-domain THz spectroscopy to study thin films of CeCoIn$_5$ in its heavy-fermion state. Below the Kondo coherence temperature $T^*$, a narrow Drude peak emerges with decreasing temperature, indicating the $f$ electron - conduction electron hybridization and formation of the heavy-fermion state. By fitting the optical conductivity with two-Drude components, the THz conductivity is connected to the dc data in a Kramers-Kronig consistent fashion. We characterize the optical scattering rate and effective mass of the heavy carriers using extended Drude model analysis. The electronic scattering rate shows a linear dependence on temperature below 20 K, consistent with the dc resistivity.  In contrast, the full renormalized quasiparticle scattering rate, which is the scattering rate of the heavy quasiparticles shows a $T^2$ dependence.  This shows the possibility of a hidden Fermi liquid phase in CeCoIn$_5$.

This project at JHU was supported by the Gordon and Betty Moore Foundation EPiQS Grant No. GBMF-9454, and NSF-DMR 2226666.  NPA had additional support from the Quantum Materials program at the Canadian Institute for Advanced Research.  At Cornell this work was supported by the Gordon and Betty Moore Foundation's EPiQS initiative through Grant Nos. GBMF3850 and GBMF9073.  Additional support for materials synthesis was provided by the National Science Foundation through Grant No. DMR-2104427 and Cooperative Agreement No. DMR-2039380 through the Platform for the Accelerated Realization, Analysis, and Discovery of Interface Materials (PARADIM).  Work at LANL was carried out under the auspices of the U.S. Department of Energy, Basic Energy Sciences, Materials Sciences and Engineering Division.  We acknowledge useful conversations with G. Kotliar.

%%  Although emphasized in both classic work on e-p interaction and in the havy fermion context. effective mass in the formula for conductivity should be the band-structure mass and not the renormalized many body mass, which occurs for example in the specific heat. 

%\begin{acknowledgments}

%\end{acknowledgments}

%\nocite{*}
%\bibliographystyle{apsrev4-2}
\bibliography{CeCoIn5_database}% Produces the bibliography via BibTeX.

\bigskip

{\it Upon completion of this work we became aware of  unpublished interpretation of backwards wave oscillators data regarding a hidden-Fermi liquid in CeCoIn$_5$ in the thesis of U. Pracht~\cite{pracht2017electrodynamics}.}

\end{document}